\documentclass[fleqn,10pt]{wlscirep}
\usepackage[utf8]{inputenc}
\usepackage[T1]{fontenc}
\usepackage{mathpazo}

\DeclareMathAlphabet{\mathcal}{OMS}{cmsy}{m}{n}

\usepackage{caption}
\usepackage{subcaption}
\usepackage{amsmath}

\title{Bayesian Optimization of Bose-Einstein Condensates}

\author[1,+,*]{Tamil Arasan Bakthavatchalam }
\author[2,+]{Suriyadeepan Ramamoorthy}
\author[2,+]{Malaikannan Sankarasubbu}
\author[3,+]{Radha Ramaswamy}
\author[1,+]{Vijayalakshmi Sethuraman}
\affil[1]{Department of Physics, Presidency College (Autonomous), University Of Madras, Chennai-600005, India}
\affil[2]{Saama AI Research Lab,Chennai-600032, India}
\affil[3]{Centre for Nonlinear Science (CeNSc), Post-Graduate and Research Department of Physics, Government College for Women (Autonomous), Kumbakonam-612001, India}

\affil[*]{tamilarasanbakthavatchalam@gmail.com}

\affil[+]{these authors contributed equally to this work}


\begin{abstract}
Machine Learning methods are emerging as faster and efficient alternatives to numerical simulation techniques. The field of Scientific Computing has started adopting these data-driven approaches to faithfully model physical phenomena using scattered, noisy observations from coarse-grained grid-based simulations. In this paper, we investigate data-driven modelling of Bose-Einstein Condensates (BECs). In particular, we use Gaussian Processes (GPs) to model the ground state wave function of BECs as a function of scattering parameters from the dimensionless Gross Pitaveskii Equation (GPE). Experimental results illustrate the ability of GPs to accurately reproduce ground state wave functions using a limited number of data points from simulations. Consistent performance across different configurations of BECs, namely Scalar and Vectorial BECs generated under different potentials, including harmonic, double well and optical lattice potentials pronounces the versatility of our method. Comparison with existing data-driven models indicates that our model achieves similar accuracy with only a small fraction ( $\frac{1}{50}$th ) of data points used by existing methods, in addition to modelling uncertainty from data. When used as a simulator post-training, our model generates ground state wave functions $36 \times $ faster than Trotter Suzuki, a numerical approximation technique that uses Imaginary time evolution. Our method is quite general; with minor changes it can be applied to similar quantum many-body problems.
\end{abstract}
\begin{document}

\flushbottom
\maketitle
%
%
\thispagestyle{empty}


\section{Introduction}
\label{Introduction}
At ultra-cold temperatures, when a gas is dilute enough such that the interactions are weak, the atoms crash into the ground state, leading to an exotic state of matter called Bose-Einstein Condensates (BECs)\cite{ketterle1999experimental}. BECs are in general governed by the mathematical model described by Gross-Pitaevskii equation (GPE), a variant of Nonlinear Schr\"{o}dinger equation (NLSE). In GPE, there are two variants, namely nonlinear interaction parameter (coupling strength) and the trapping potential and both could be spatio-temporal in general. GPE in its most general form cannot be solved analytically. It can be solved analytically\cite{radha2015analytical} only for specific choices of coupling strengths and trapping potentials. In other cases, one has to resort to numerical techniques \cite{baomathematical} like Crank-Nicholson Scheme, Trotter-Suzukii Approximation, Finite Element Analysis, etc.

Numerical techniques rely on numerical differentiation which is the finite difference approximation of derivatives using values of the original function evaluated at some sample points. Numerical approximations of derivatives are inherently ill-conditioned and unstable due to introduction of truncation and round-off errors \cite{ad4ml}. On the other hand, Automatic Differentiation (AD) also called algorithmic differentiation is a family of techniques for efficiently and accurately evaluating derivatives of numeric functions expressed as computer programs \cite{ad}. AD exploits the fact that every computer program executes a sequence of elementary arithmetic operations and elementary functions. By applying the chain rule repeatedly to these operations, derivatives of arbitrary order can be computed automatically and accurately to working precision. AD is both efficient and numerically stable compared to Numerical Differentiation. AD has become the beating heart of Modern Machine Learning, leading to a new paradigm of programming called Differentiable Programming.

In recent years, Scientific Computing community has started adopting data-driven approaches from Machine Learning, as faster and efficient alternatives to numerical simulation techniques. This new regime called Data-driven Scientific Computing has shown great promise already by faithfully modelling physical phenomena using scattered noisy observations from coarse-grained grid-based numerical simulations. The learned models are capable of predicting the dynamics of the system under study in a fixed number of CPU cycles irrespective of the dimensions of the grid. Unlike numerical techniques that rely on unstable numerical gradients, Machine Learning models depend on precise and stable automatic differentiation.

Advances in Machine Learning has led to incredible breakthroughs in areas such as Computer Vision\cite{efficientdet}, Natural Language Processing \cite{fbwmt}, Computational Biology \cite{insilico}, etc, demonstrating super-human abilities \cite{planet, poker, atari, alphago} in certain tasks. These dramatic improvements are reflected in computational sciences that make use of Machine Learning. Artificial Neural Networks (ANNs), the focal point of Modern Machine Learning or Deep Learning, are powerful and versatile models that can automatically learn representations from data. ANNs were primarily used in Physical Sciences, to learn suitable order parameters and detect the phase transition, without any prior domain knowledge \cite{carrasquilla2017machine,morningstar2017deep,tanaka2017detection,zdeborova2017machine}.   Greitemann et al. \cite{greitemann2019identification} used a kernel-based learning method to learn the phases in magnetic materials to identify complex order parameters. In the context of nonlinear dynamical systems, Jaegar et al. \cite{jaeger2004harnessing} employed ANNs to predict the trajectories of a chaotic time-series improving the accuracy by a factor of 2400 over previous techniques. ANNs were used as a variational representation of quantum states in quantum-physical many-body problems to cope up with the exponential complexity of the many-body wave function \cite{carleo2017solving}.  Nomura et al. \cite{PhysRevB.96.205152} employed ANNs to develop a machine learning method for constructing accurate ground-state wave functions of strongly interacting and entangled quantum spin systems as well as fermionic models on lattices. A restricted Boltzmann machine algorithm in the form of an ANN, combined with a conventional variational Monte Carlo method turned out to be a highly accurate quantum many-body solver capable of accurately predicting ground-state wave functions of strongly interacting and entangled quantum spin systems as well as fermionic models on lattices \cite{gao2017efficient}. ANNs have been proven to be capable of representing large-scale quantum states, including the ground states of many-body Hamiltonians and states generated by quantum dynamics \cite{carleo2018constructing,czischek2018quenches, schmitt2018quantum, fabiani2019investigating}.

Despite being robust and flexible, ANNs suffer from a  number of shortcomings. ANNs are data-hungry and hence they do not play well with small data. They are over-confident in predictions and are susceptible to adversarial inputs. Learning in ANNs is based on Maximum Likelihood Estimation which leads to point estimates of learnable parameters instead of probability distributions. Gaussian Processes (GPs) on the other hand, are flexible probabilistic models that can learn probability distributions over functional mapping of inputs to outputs. They are quite efficient in learning hidden dynamics from small data. Being a Bayesian method, GP inherently models uncertainty from data. Learning results in a posterior distribution from which we could draw new samples. 

 Even though Machine Learning approach has been employed in modelling BECs earlier by Liang etal\cite{liang}, they exploited Convolutional Neural Network architecture(CNN) to model the ground state wave functions of BECs in different platforms. Liang et al. \cite{liang} showed that it is possible to approximate the wave function using a neural network but do not guarantee any kind of improvement over conventional numerical techniques\cite{baonumerical}. Our investigation bridges this gap by using a Gaussian Process as a cheaper, powerful and efficient alternative to neural networks in the context of modelling ground state wave functions in BECs.


In this paper, our goal is to examine whether Gaussian Processes could be a viable surrogate model for data-driven emulation of one-dimensional Scalar and Vectorial BECs. Our validation of GP is centred around the following aspects:

\begin{enumerate}
\item \textit{Correctness}: Can GP accurately model the ground state of BECs ?
\item \textit{Versatility}: Can GP adapt to different settings of BECs?
\item \textit{Data Efficiency}: How many data points are necessary to model a wave function?
\item \textit{Compute Efficiency}: How do they fare against numerical techniques when used as a simulator, after training?
\end{enumerate}

To answer these questions, we studied the performance of Gaussian Processes on different settings of BECs. We employ Trotter-suzuki approximation, a numerical technique for simulating the ground state wave functions, $\psi$. We run simulations by setting up a one-dimensional grid and vary the coupling strength $g$, the parameter that controls the interaction between the atoms besides varying the trapping potential. Similarly for two-component BECs, we run simulations by varying interaction parameters $\{g_{11}, g_{12}, g_{22}\}$ which results in two wave functions $\{\psi_1, \psi_2\}$. We model the simulated wave functions using a GP with Radial Basis Function (RBF) kernel\cite{rbf}. The GP models the wave function as a function of space $x$ and coupling strength $g$. The results reveal the versatility of GP in modelling different kinds of wave functions, efficiently and accurately, with uncertainty estimates. In addition to modelling uncertainty from data, our method performs better than Liang et al. \cite{liang} in terms of efficiency and model complexity. Our model being simpler in terms of model complexity, uses just a small fraction ($\frac{1}{50}$th) of the data points used by Liang et al. to achieve similar accuracy. Furthermore, comparing the efficiency of our method in predicting wave functions, with Trotter-suzuki approximation, we find that our method performs $36 \times$ faster.

The paper is organised as follows: Section.\ref{Background} lays the foundation for the rest of the paper. It starts with a brief introduction to Bose-Einstein Condensates in Section.\ref{eg1}, Trotter-Suzuki Approximation in Section.\ref{eg2}, and a detailed description of Gaussian Processes in \ref{eg3}. Section.\ref{Methods} provides the reader with the tools necessary to replicate our experiments. Section.\ref{Experiments} includes a report on all our experiments using Gaussian Processes as a surrogate model for different settings of BECs. Section.\ref{Discussion} discusses the implications of our experimental results and summarizes possible avenues for further research.



\section{Background} 
 \label{Background}

\subsection{Bose-Einstein Condensates}
\label{eg1}

At temperatures close to absolute zero, majority of the bosons in a gas crash into the ground state of the system creating an exotic state of matter known as Bose-Einstein Condensates (BECs). The dynamics of BECs is described by Gross-Pitaevskii \cite{rogel2013gross} equation \ref{eq1} which belongs to the family of Variable Coefficient Nonlinear Schr\"{o}dinger equations, given by

\begin{equation}\label{eq1}
i \hbar \frac{\partial\psi}{\partial t} = (-\frac{\hbar^{2}}{2 m } \triangledown^{2} + V_{ext} + g |\psi|^{2})\psi
\end{equation}

where $\psi$ denotes the wave function(or order parameter) of BECs, $m$ is mass the atoms of the condensate, $g$ is the inter-atomic interaction, $V_{ext}$  is the trapping potential and $|\psi|^{2}$ is the atomic density. 

A two component BEC known as Vector BEC is  endowed with inter-atomic and intra atomic interaction in addition to the trap and hence its density can be manipulated with more freedom and is governed by Coupled Gross-Pitaveskii Equation \cite{baomathematical} of the form equation \ref{eq2} and  \ref{eq3},

\begin{equation}\label{eq2}
i \hbar \frac{\partial\psi_{1}}{\partial t} = (-\frac{\hbar^{2}}{2 m } \triangledown^{2} + V_{ext}(x) + g_{11} |\psi_{1}|^{2}+ g_{12} |\psi_{2}|^{2})\psi_{1}
\end{equation}

\begin{equation}\label{eq3}
i \hbar \frac{\partial\psi_{2}}{\partial t} = (-\frac{\hbar^{2}}{2 m } \triangledown^{2} + V_{ext}(x) + g_{21} |\psi_{1}|^{2}+ g_{22} |\psi_{2}|^{2})\psi_{2}
\end{equation}

\subsection{Trotter-Suzuki Approximation}
\label{eg2}
The behaviour of any physical system can be studied by solving the partial differential equations (PDEs) which represent the dynamics of that physical phenomenon. In practice, most PDEs with any real application are nonlinear in nature and are hard to solve analytically. This is particularly true for complex dynamical systems which are quite difficult to solve and necessitate high computational resources to arrive at highly accurate solutions. Trotter-suzuki decomposition implemented by Wittek and Cucchietti \cite{peterwittek}, exploits optimized kernels to solve Gross-Pitaevskii equation of a free particle. The exponential operators in PDEs are notoriously hard to approximate. Trotter Suzuki decomposes the Hamiltonian into sum of diagonal matrices which eases the task of computing the exponential. The evolution operator is calculated using the Trotter-Suzuki approximation. Given an Hamiltonian as a sum of hermitian operators, for instance $H=H_1+H_2+H_3$, the evolution is approximated as \cite{peterwittek}
\begin{equation}
 e^{-i\Delta tH} = e^{-i\frac{\Delta t}{2} H_1} e^{-i\frac{\Delta t}{2} H_2} e^{-i\frac{\Delta t}{2} H_3} e^{-i\frac{\Delta t}{2} H_3} e^{-i\frac{\Delta t}{2} H_2} e^{-i\frac{\Delta t}{2} H_1}.
 \end{equation} 
\subsection{Gaussian Processes}
\label{eg3}
Gaussian Processes \cite{humblegaussian} are probabilistic machine learning models that define a distribution over functions. They are infinite dimensional realizations of a multi-variate Gaussian Distribution. Given a set of points in the input space $\{x_1, x_2, .. x_n\}$ and the function $f$ evaluated at those points $\{f^{1}_{e}, f^{2}_{e}, ..., f^{n}_{e} \}$, we can formally define a GP as:

\bigbreak  

\textbf{Definition} : $p(f)$ is a Gaussian Process if for any subset $\{x_1, x_2, ... x_n\}  \subset \mathcal{X}$, the marginal distribution over the subset $p(f_{e})$ has a multivariate Gaussian distribution.

\bigbreak  

\begin{figure}
\begin{subfigure}{.5\textwidth}
  \centering
  \includegraphics[width=1\linewidth]{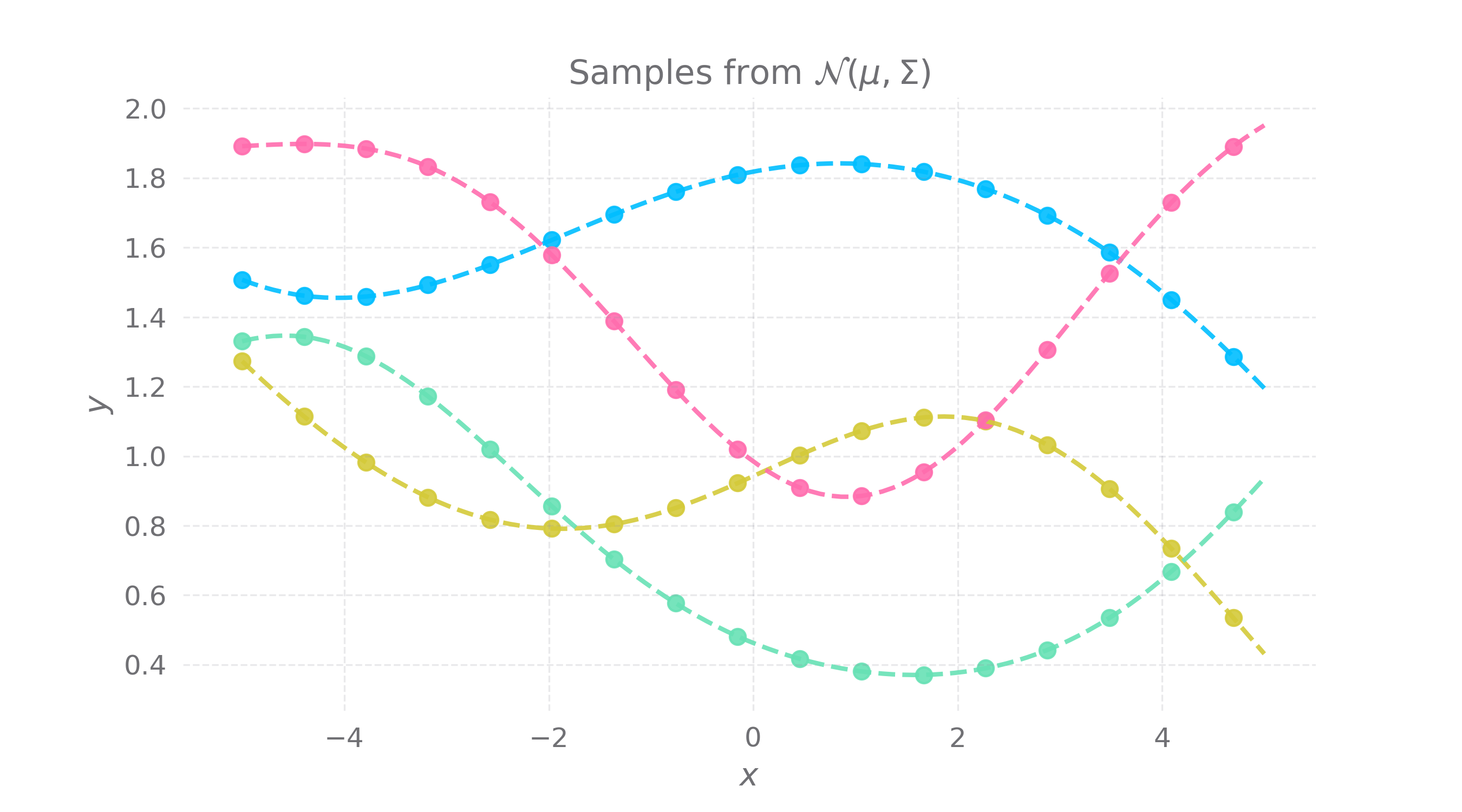}
  \caption{}
  \label{figure_1_a}
\end{subfigure}%
\begin{subfigure}{.36\textwidth}
  \centering
  \includegraphics[width=1\linewidth]{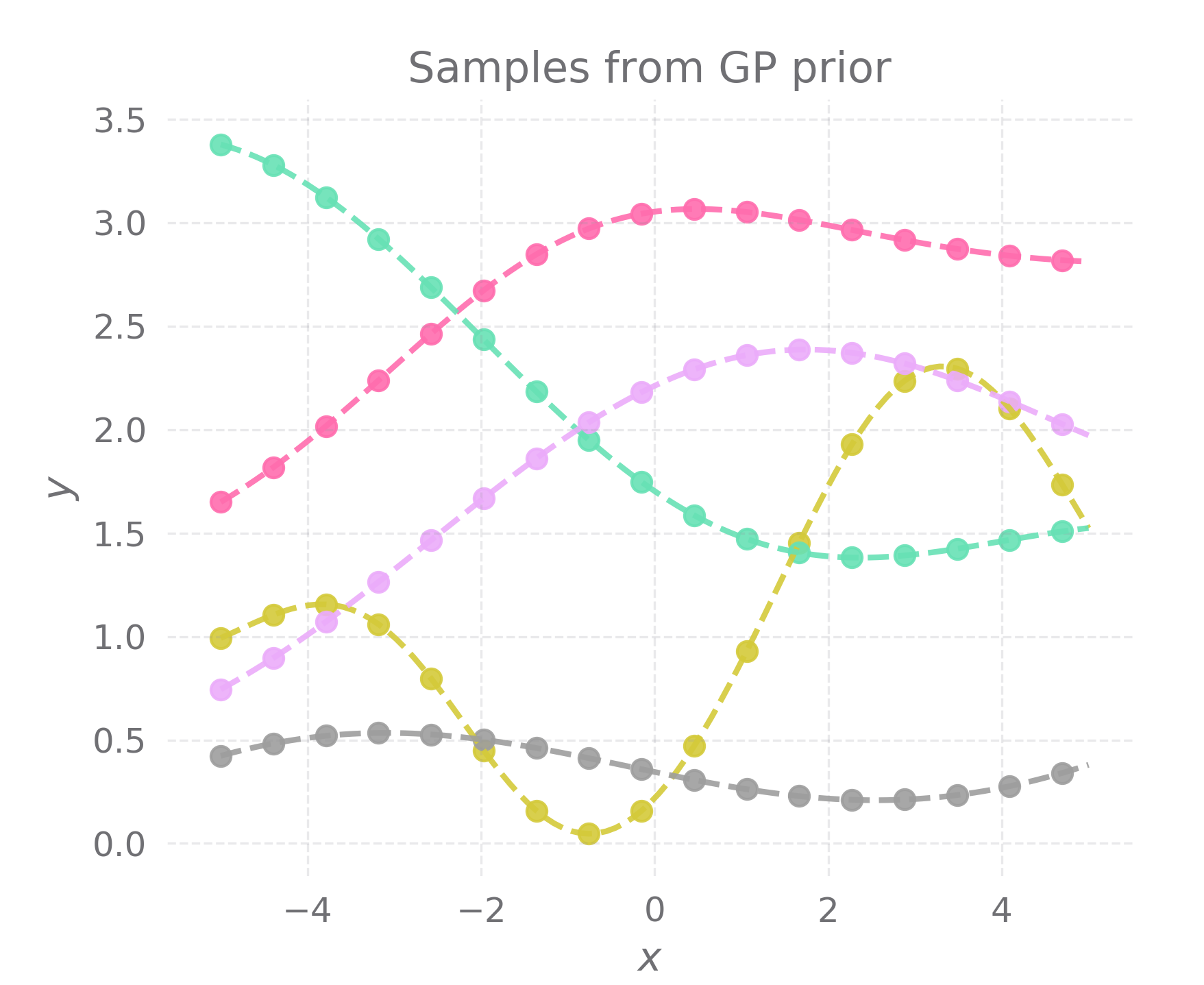}
  \caption{}
  \label{figure_1_b}
\end{subfigure}
\caption{
\textbf{(a)} Four samples from a multi-variate normal distribution $\mathcal{N}(0, \Sigma)$ plotted sequentially.
\textbf{(b)} Samples from GP prior conditioned on different values of length scale $l_k$.
}
\label{fig:gp_sample}
\end{figure}

Consider a Multivariate Gaussian Distribution given by,

										$$z_i \sim \mathcal{N}(\mu, \Sigma)$$

where $\mu \in \mathbb{R}^k$ is a $k$-dimensional zero vector and $\Sigma \in \mathbb{R}^{k \times k}$ is the covariance matrix that captures the correlation between dimensions $\{X_1, X_2, .. X_{k}\}$. Let us sample $4$ points $\{z_1, z_2, z_3, z_4 \}$ from the distribution and plot it sequentially as shown in Figure.\ref{figure_1_a}. From Figure.\ref{figure_1_a}, we can make two inferences. (1) Points closer to each other on the x-axis behave similarly on the y-axis (2) Points farther apart from each other behave differently on the y-axis. That is, closer points are highly correlated with each other while father points are not. This intuition leads to the understanding that by connecting these points sequentially, we can realize smooth functions like the ones shown in Figure.\ref{figure_1_b}. The functions we realized by connecting the points sequentially can be considered as samples from a Gaussian Process  by a zero-mean vector $\mu$ and the hitherto unknown covariance matrix $\Sigma$.

Yet another inference that can be made from figure.\ref{figure_1_a} is that the nature of the functions realized are dependent on the covariance matrix $\Sigma$. By manipulating the covariance matrix $\Sigma$, one can manipulate the kind of functions realized from $k$-dimensional samples $\{z_i\}$. The mechanism to manipulate the covariance matrix is through an algebraic function named \textit{Covariance function} or \textit{kernel}. The kernel $k(x, x') \rightarrow \mathcal{R}$  takes two points from the input space $\mathcal{X}$ as inputs and returns a scalar value $c \in [0, 1]$. This value represents the similarity between the two input points. The covariance matrix $\Sigma$ is constructed by calculating the similarity between $k$ equally spaced points in the input space $\{x_1, ... x_k\}$. By controlling the kernel function, we control the covariance matrix that governs the GP, thereby controlling the nature of functions generated. The kernel is usually parameterized by one or more tunable variables which act as knobs for controlling the shape of the functions that GP generates. 

Let us assume for didactic purposes that the $x$ and $y$ are scalars and the function $f$ is given by $f : \mathcal{X} \rightarrow R$. We consider Radial Basis Function (RBF) parameterized by length scale $k_l$ as the kernel. 
\begin{equation}\label{eq5}
k(x_i, x_j) = exp \Big{(} -\frac{1}{2 k_l^2} (x_i - x_j)^2 \Big{)}
\end{equation}
Length scale $k_l$ controls the order of the functions generated as evident from Figure.\ref{figure_1_b}. 

\subsubsection*{Inference}
Inference in GP is a two-step process : \textit{Kernel Parameter Search} and \textit{Posterior Estimation}. In kernel parameter search, we learn the kernel parameter values that fit the data well. The optimal value of kernel parameters are estimated using Maximum Likelihood Estimation (MLE) by minimizing marginal negative log likelihood \cite{gpml} given by

\begin{equation}\label{eq6}
log\ p(y|X) = log\ \mathcal{N}(y | 0, K_y) =
-\frac{1}{2} y^T K_y^{-1} y\ -\frac{1}{2}log\ |K_y|\ -\frac{N}{2}log\ (2\pi)
\end{equation}
\begin{equation}
         K_y = k(X, X)
\end{equation}
where $N$ is the number of data points, $(X, y)$ represent the data points.

Then, we estimate GP posterior conditioned on data i.e. mean vector and covariance matrix conditioned on data. The definition of GP states that the joint distribution of observed data $y$ and predictions $f_*$ has a multivariate Gaussian distribution. Given data $X$ and new input $X_{*}$, we can write the joint distribution as,
\begin{equation}
\begin{bmatrix}
           y \\
           f_{*}
         \end{bmatrix} \sim
         \mathcal{N}
         \begin{pmatrix} 
         \textbf{0},
				 \begin{pmatrix} 
         K_y & K_{*} \\ 
         K_{*}^T & K_{**}  
         \end{pmatrix}
         \end{pmatrix}
        \end{equation}
         
\begin{equation*} K_{*} = k(X, X_{*})\\ \end{equation*}
\begin{equation*}K_{**} = k(X_{*}, X_{*}) \end{equation*}

Using the Multivariate Gaussian Theorem \cite{kevinmurphy}, we arrive at the parameters of the posterior, 
$$\mu_{*} = K_{*}^T K_y^{-1}y$$
$$\Sigma_{*} = K_{**} - K^T_{*}K_y^{-1}K_{*}$$

\subsubsection*{Motivation}

In our experiments, we consider the ground state wave function of one-dimensional BECs\cite{baonumerical}. The simulation takes as input multiple parameters including grid parameters like radius, length, potential function ($V_{ext}$), coupling strength ($g$),etc., and generates a one-dimensional wave function using Imaginary Time Evolution method. The time period to complete a simulation depends on the dimensions and size of the grid. The promise of Data-driven Scientific computing is the ability to model any process without closed-form solutions using experimental data and building a predictive model that can provide predictions at any point in the grid with absolute guarantee. As discussed in the Introduction, a data-driven model trained on scattered data points from simulations is capable of predicting the wave function at any point within the grid in a few fixed number of CPU cycles irrespective of grid dimensions or size.

Gaussian Processes model the wave function as a probability distribution over functions supported by a kernel with optimal parameters. Being a Bayesian model, GP is data-efficient and makes uncertainty-aware predictions. The ability to model uncertainty is especially important in modelling physical phenomenon from a few scattered noisy observations. Bayesian methods rely on Bayesian Inference which estimates posterior over unknowns in contrast to non-probabilistic models which result in point estimates of unknowns. As a consequence, it is possible to generate entire wave functions from the GP that are subject to constraints on the input parameters like coupling strength, omega(Rabi Coupling), etc.


\section{Methods}
\label{Methods}

We use trotter-suzuki-mpi \cite{tsmpi}, a massively parallel implementation of the Trotter-Suzuki approximation to simulate the evolution of Bose-Einstein Condensates. The simulation setup consists of a one-dimensional grid which represents the physical system. This discretized space is defined by a lattice of 512 nodes within a physical length of 24 units. The physics of BECs is described by the Hamiltonian which represents the GPE equation. The Hamiltonian requires a grid and a trapping potential. The trapping potential is defined as a function which takes $x$ and $y$ as arguments and returns a scalar value as output. In this case, we consider a harmonic trapping potential given by $\frac{x^2}{2}$. The state of the system is initialized in gaussian form.  Imaginary time evolution evolves the system state using the dynamics defined by the Hamiltonian, in $10^{3}$ iterations with time step being $\Delta_t = 10^{-4}$ for each iteration. In order to collect data, we run $M$ simulations by varying the value of the interaction parameter $g$. We use sklearn's Gaussian Process API \cite{sklearn} with Radial Basis Function (RBF) kernel as a surrogate to model the wave function $\psi$ as a simpler but continuous function of space $x$ and the interaction parameter $g$. 

The prior on the groundstate wave function of BEC is given by 

$$\psi_{prior} \sim GP(0, k(x, x'; \theta))$$

$N$ data points of the form $(x, g, \psi)$ are sampled randomly from $M$ simulation outcomes. Throughout our experiments, we use the RBF kernel given in equation.\ref{eq5}.

 The hyper-parameters $\theta$ of the kernel are tuned using Maximum Likelihood Estimation which essentially amounts to finding the parameters that minimize the expression in equation.\ref{eq6}, given a list of data points.

\begin{figure}
\begin{subfigure}{.42\textwidth}
  \centering
  \includegraphics[width=1\linewidth]{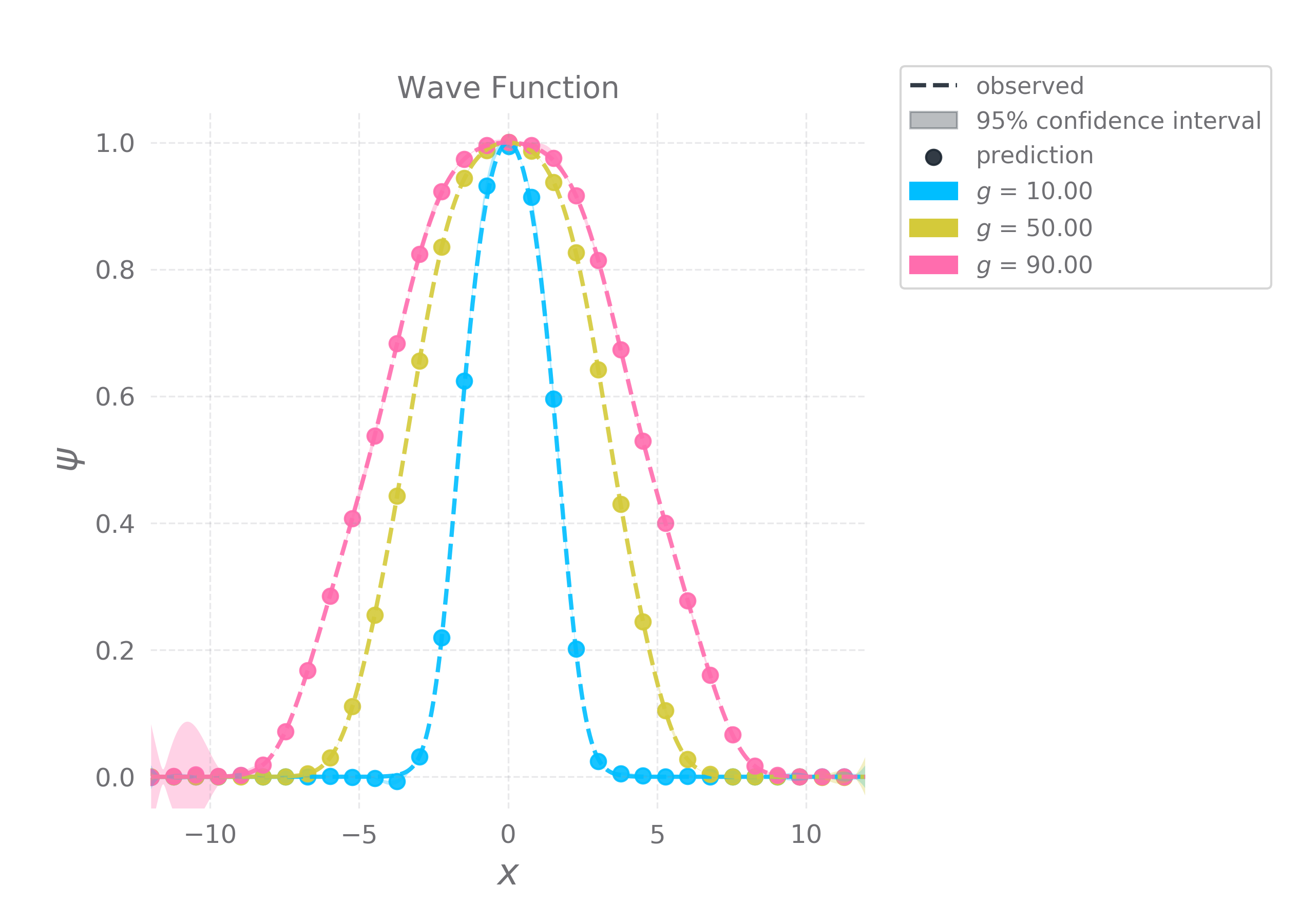}
  \caption{}
  \label{figure_2_a}
\end{subfigure}%
\begin{subfigure}{.5\textwidth}
  \centering
  \includegraphics[width=1\linewidth]{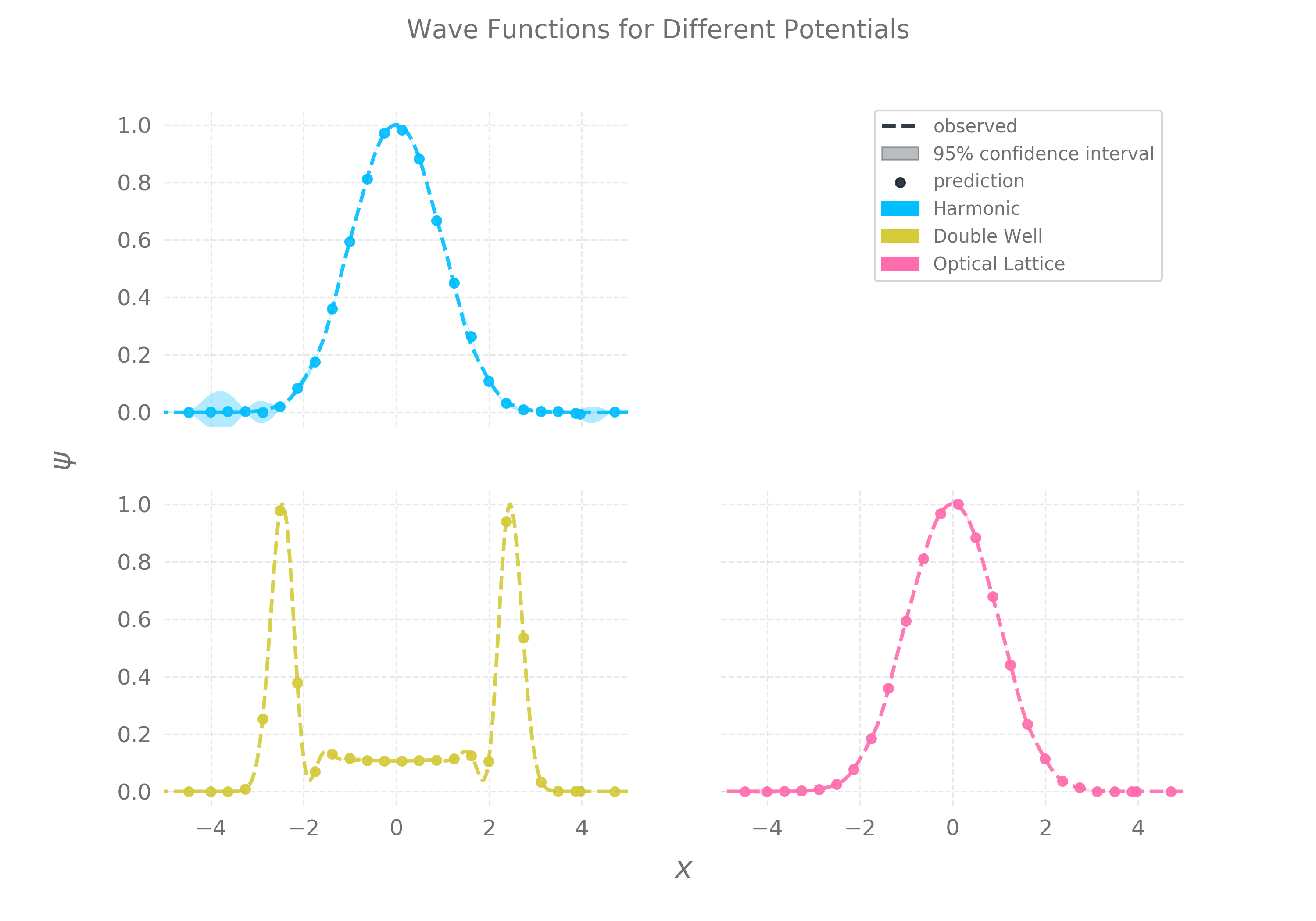}
  \caption{}
  \label{figure_2_b}
\end{subfigure}
\caption{
\textbf{(a)} Simulated and predicted ground state wave functions for different values of the interaction parameter $g = {10, 50, 90}$
\textbf{(b)} Prediction of wave functions conditioned on different trapping potentials - Harmonic, Optical Lattice, Double Well potentials.
}
\label{fig:harmonic}
\end{figure}

 The RBF kernel consists of the following parameters: length scale $k_{l}$ and variance $k_{\sigma}$. $\psi_{posterior}$ is estimated by conditioning $\psi_{prior}$ on the sampled data points. The expression for $\psi_{posterior}$ reduces to a mathematically tractable form given by the mean vector and covariance matrix presented in equation 8.
  Inference consists of calculating the marginal mean $\mu_{2|1}$ and covariance $\Sigma_{2|1}$ by substituting the training points $X$ and new test points $X_*$ into the analytical forms presented in equation 8. The marginal mean $\mu_{2|1}$ constitutes the predicted wave function. The diagonal elements of the covariance matrix form the variance $\sigma$ on each prediction. Together, they make the GP posterior parameterized by $[\mu_{2|1}, \Sigma_{2|1}]$. 
  The fidelity of predicted wave function is measured by calculating Mean Squared Error (MSE) metric against the ground truth data. Figure.\ref{fig:harmonic} portrays the results of an illustrative experiment on one-dimensional BECs with a harmonic trapping potential.

The experimental setup described above can be extended to different settings of simulation and data-driven approximation, to include two-component BECs and different kinds of trapping potentials. A detailed report of the experiments we conducted is presented in the next section.

\section{Experiments}
\label{Experiments}
All the experiments to be covered in this section will follow along similar lines as the illustrative experiment discussed in the previous section: Simulation, Data-driven Approximation and Evaluation. In the previous experiment, we ran $300$ simulations shown in Figure.\ref{figure_2_a} for equally spaced values of $g$ ranging from $1$ to $100$. We randomly sampled $500$ samples of the form $(x, g, \psi)$ from the simulation results, which we used to fit the GP. Evaluation leads to a MSE score of $1.23 \times 10^{-7}$ which indicates that the trained GP can accurately predict ground state wave functions for $100$ different values of the interaction parameter $g$ within the range $(1, 100)$.

\begin{figure} 
  \centering
  \includegraphics[width=1\linewidth]{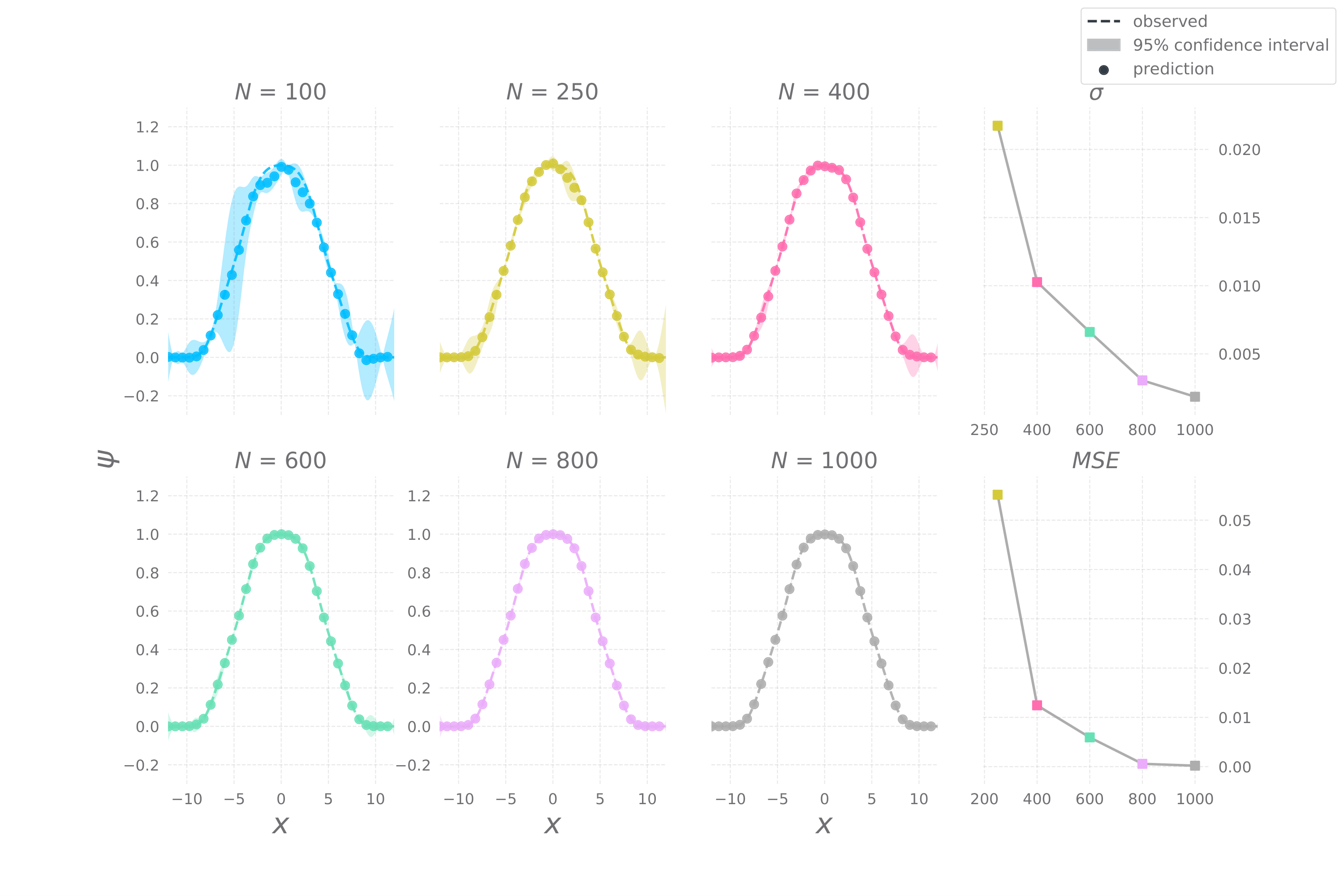}
  \caption{Gaussian Processes fit on training samples of different sizes, $N = \{100, 250, 400, 600, 800, 1000\}$Variance $\sigma$ and Mean-squared Error (MSE) plotted against training sample size.}
  \label{figure_3}
\end{figure}

To follow up, we ask the question, How many data points are necessary to build a reliable predictive model of ground state wave function? In order to answer this question, we vary the number of training samples and observe the effect it has on MSE score of GP trained on those samples. We reuse the experimental setup from the previous experiment and vary the number of data points $N$ sampled from $M=300$ simulations. The results are presented in Figure.\ref{figure_3}. Figure.\ref{figure_3} depicts the relationship between variance and number of samples $N$ and it  shows the rate of decrease of MSE w.r.t $N$. The MSE curve shows a trend of saturation close to $0$ as the number of samples are increased as shown in Table.\ref{tab:error}. Decrease in variance and MSE is evidence of confidence and accuracy in prediction respectively.

\begin{table}
  \caption{Error Estimates from Experiments}
    \label{tab:error}
\begin{tabular}{lcc}
      \textbf{Experiment} & \textbf{Number of training samples} & \textbf{Mean Squared Error}  \\  \hline
      One-component BEC with Harmonic Potential & 500 & $1.23 \times 10^{-7}$  \\  \hline
      One-component BEC with Double Well Potential & 500 & $1.57 \times 10^{-6}$    \\  \hline
      One-component BEC with Optical Lattice Potential & 500 & $4.30 \times 10^{-5}$   \\  \hline
      Two-component BEC with Harmonic Potential & 1000 & $2.52 \times 10^{-5}$   \\  \hline
      Two-component BEC with Double Well Potential & 1000 & $2.43 \times 10^{-4}$   \\  \hline
      Two-component BEC with Optical Lattice Potential & 1000 & $3.27 \times 10^{-4}$   \\  \hline
      Rabi-coupled Two-component BEC with Cosine Potential & 1200 & $4.20 \times 10^{-4}$   \\ \hline
\end{tabular}
\end{table}

\begin{figure}
\begin{subfigure}{.5\textwidth}
  \centering
  \includegraphics[width=1\linewidth]{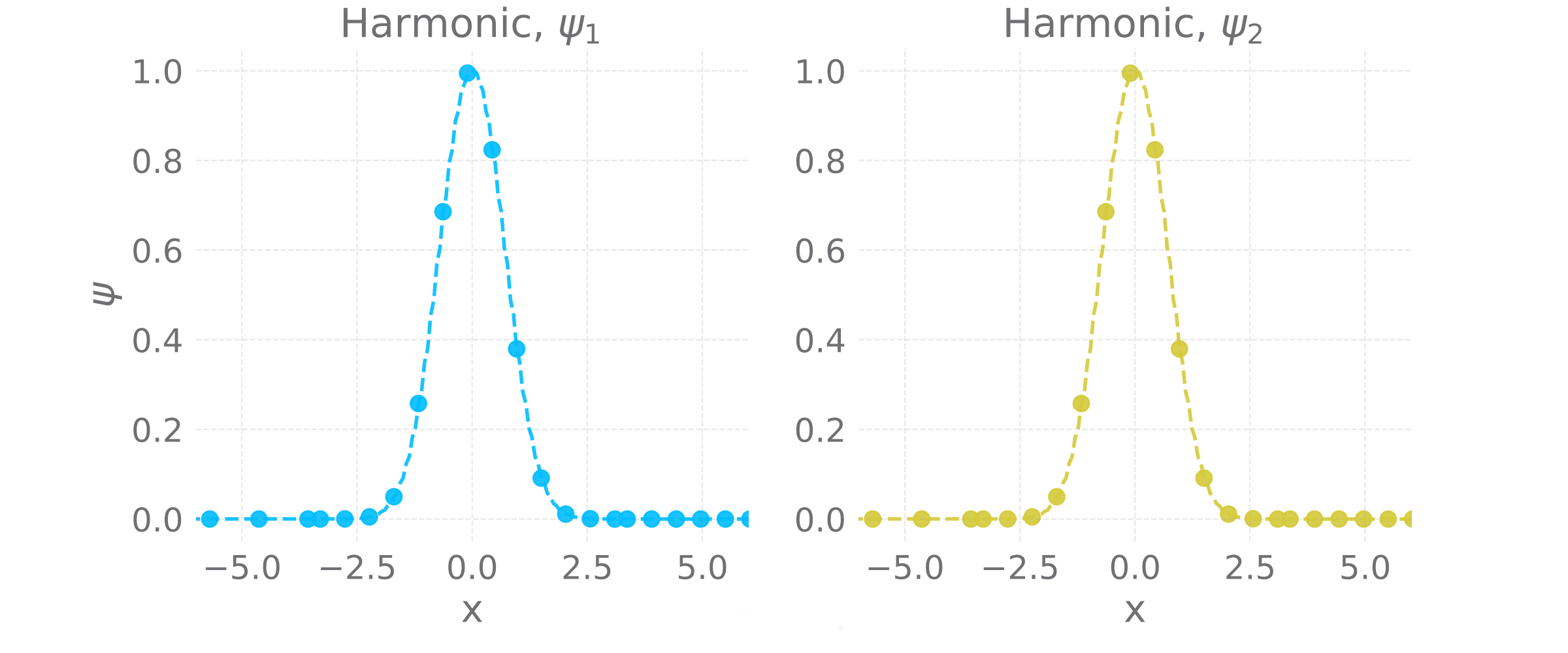}
  \caption{}
  \label{harm_double}
    \includegraphics[width=1\linewidth]{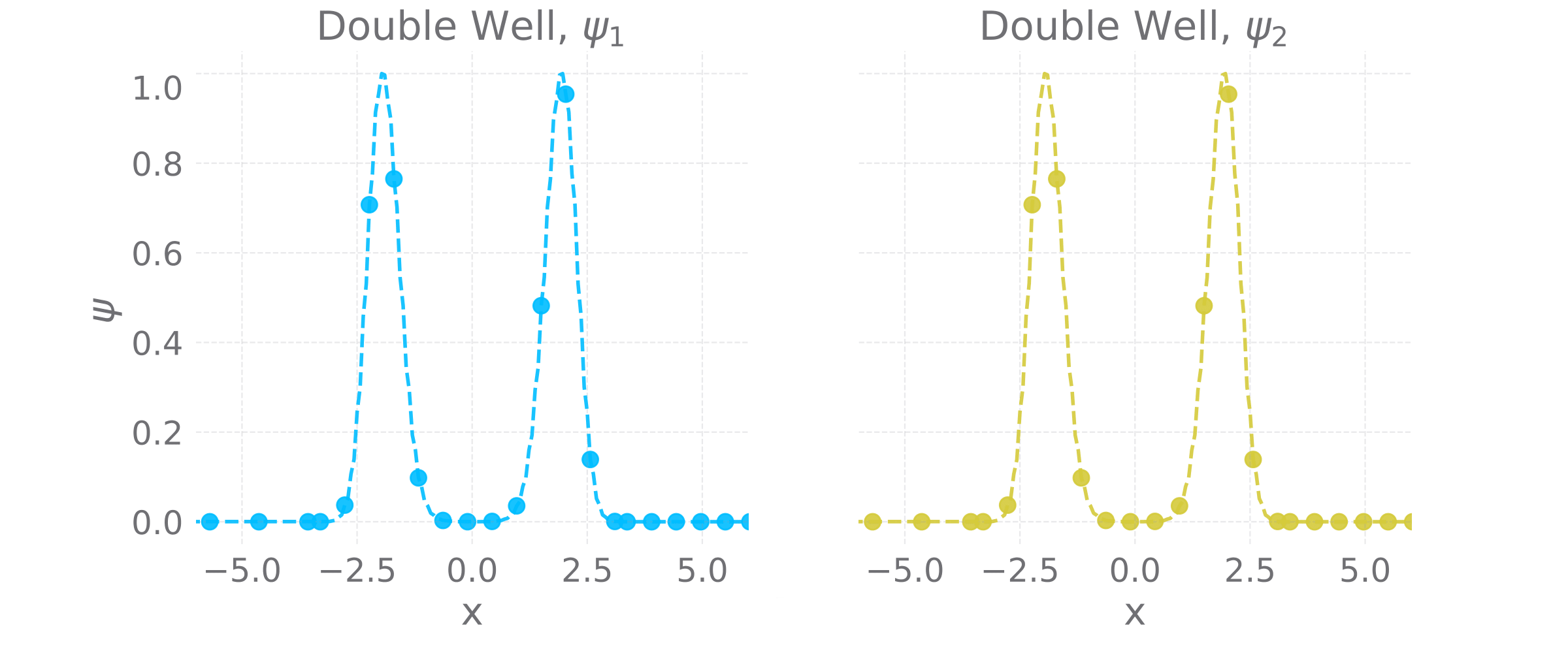}
    \caption{}
  \label{sfig:e4}
\end{subfigure}%
\begin{subfigure}{.5\textwidth}
  \centering
  \includegraphics[width=1\linewidth]{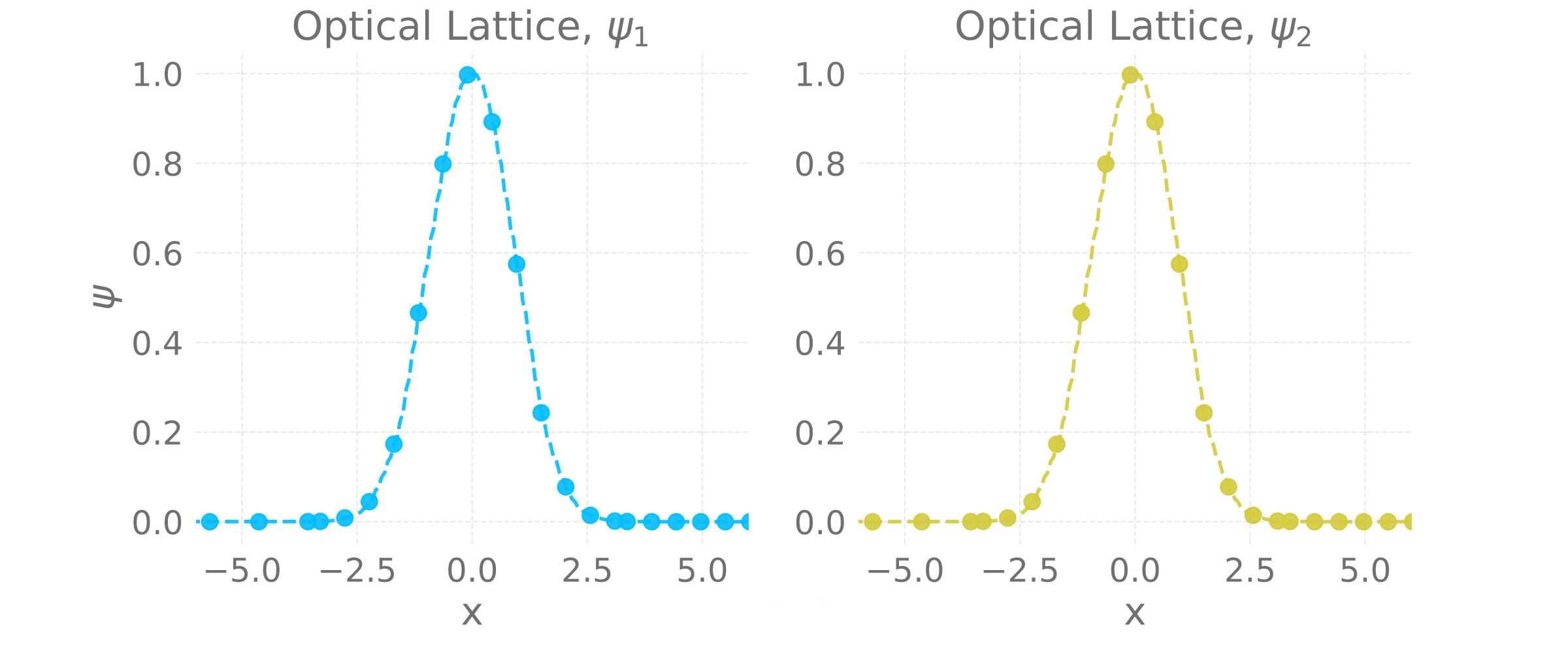}
  \caption{}
  \label{figure_4_c}
    \includegraphics[width=1\linewidth]{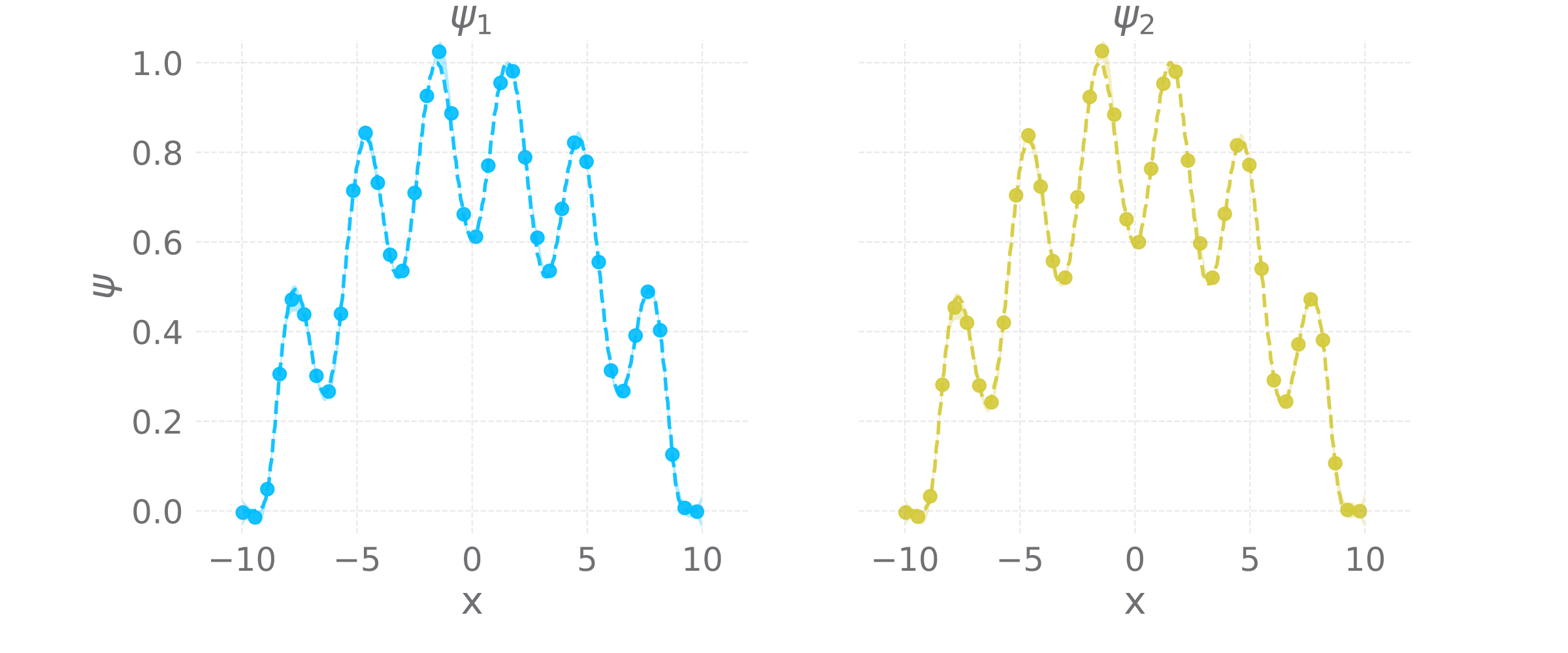}
    \caption{}
  \label{figure_4_d}
\end{subfigure}
\caption{
\textbf{(a, b, c)} Prediction of two-component BECs ground-state wave functions for different trapping potentials \textbf{(a)} Harmonic, \textbf{(b)} Double Well, \textbf{(c)} Optical Lattice
\textbf{(d)} Emulation of Rabi-coupled two-component BECs conditioned on cosine potential 
}
\label{fig:twocomponent}
\end{figure}

The next experiment tests the versatility of GP by modelling the ground states of BECs for different choices of trapping potentials. We consider Harmonic , Double Well  and Optical Lattice potentials in Figure.\ref{fig:twocomponent} . We restrict the range of $g$ to $(0, 2)$ in order to ensure the stability of generated wave function for different potentials. We run $100$ simulations and collect $500$ samples for each potential. We fit a GP for each potential using collected data points. Evaluating the models results in MSE scores of $1.23 \times 10^{-7}$, $1.57 \times 10^{-6}$ and  $4.30 \times 10^{-5}$, for harmonic, double well and optical lattice potentials \cite{baomathematical} respectively. The predictive wave functions for a value of $g=1$ is plotted against the ground truth in Figure.\ref{fig:twocomponent}{\color{blue}(a,b,c)}.

The ground state wave functions of two-component BECs are given by $\psi_1$ and $\psi_2$. The interaction strength is defined by parameters $g_{11}$, $g_{12}$ and $g_{22}$. We model the system using a GP which learns the relationship between $(\psi_1, \psi_2)$ and the interaction parameters. We sample $500$ data points from $200$ simulations with values of interaction parameters lying within the range $(-1, 1)$. The choice of this range is dictated by the zone of stability of the system. Predictions and ground truth wave functions for different potentials conditioned on values of interaction parameters given by $\{g_{11}=0.1, g_{12}=-0.1, g_{22}=0.1\}$, are presented in Figure.\ref{fig:twocomponent}{\color{blue}(a,b,c)}. Evaluation on a test set within the range $g \in (-1, 1)$ results in MSE in the range of $10^{-4}$.

\begin{table}
  \caption{Trotter-Suzuki vs Gaussian Processes}
    \label{tab:time}
\begin{tabular}{lcc}
      \textbf{Setting} & \textbf{Trotter-Suzuki (ms)} & \textbf{Gaussian Processes (ms)}  \\  \hline
      One-component BEC with Grid Size $d=512$      & $160$   &  $11$       \\  \hline
      One-component BEC with Grid Size $d=1024$    & $330$  &  $22.6$    \\  \hline
      Two-component BEC with Grid Size $d=512$      & $930$  &  $14$         \\  \hline
      Two-component BEC with Grid Size $d=1024$    & $1150$ &  $23$         \\  \hline      
\end{tabular}
\end{table}

\begin{equation}\label{eqrabi}
i \hbar \frac{\partial\psi_{1}}{\partial t} = (-\frac{\hbar^{2}}{2 m } \triangledown^{2} + V_{ext}(x) + g_{11} |\psi_{1}|^{2}+ g_{12} |\psi_{2}|^{2})\psi_{1} +\Omega \psi_{2} \\
\end{equation}
\begin{equation}\label{eqrabi1}
i \hbar \frac{\partial\psi_{2}}{\partial t} = (-\frac{\hbar^{2}}{2 m } \triangledown^{2} + V_{ext}(x) + g_{21} |\psi_{1}|^{2}+ g_{22} |\psi_{2}|^{2})\psi_{2} + \Omega \psi_{1}
\end{equation}

We reproduce an experiment conducted by Liang et al. \cite{liang} using GP - Emulation of rabi-coupled two-component BECs. Figure.\ref{fig:twocomponent}{\color{blue}(d)} shows the predicted ground state wave function of rabi-coupled two-component BECs equation. \ref{eqrabi} \& \ref{eqrabi1}, conditioned on a cosine potential given by $\frac{x^2}{2} + 24cos^2 x$ with interaction parameters $\{g_{11}=103, g_{12}=100, g_{22}=97\}$. Finally, we use trained GPs as simulators to generate wave functions and compare their performance against Trotter-suzuki. We experiment with one-component and two-component BECs in $512$ and $1024$ grid spaces. The results tabulated in Table.\ref{tab:time} indicate an average of $36 \times$ speed up in wave generation with reasonable accuracy.

\section{Discussion}
\label{Discussion}
CNN-based model employed by liang et al. \cite{liang} learns a mapping from coupling strength $g$ to wave function values at fixed points in space, a scalar to vector mapping. This results in a highly limited predictive model of wave functions. In contrast, our method models the wave function as a continuous function of space and coupling strength, resulting in a predictive model which is well-defined in all the points within the chosen interval. The mean-squared error (MSE) presented by Liang et al. \cite{liang}, on an average ranges from $10^{-4}$ to $10^{-5}$. Our experimental results tabulated in Table.\ref{tab:error} show a similar MSE score ranging from $10^{-4}$ to $10^{-7}$. We use a maximum of $1000$ data points from simulations to achieve these results while Liang et al. \cite{liang} uses data from $50000$ simulations to train their CNN. Our method, only using a small fraction ($\frac{1}{50}$) of data points used by Liang et al. \cite{liang}, has demonstrated comparable accuracy in most cases and enhanced accuracy in others. Our experiments with various trapping potentials show that it is possible to model different kinds of ground states of BECs using a vanilla GP with an RBF kernel.

We compare the predictive power of our model with Trotter-suzuki Approximation - a finite difference based numerical technique. Trotter-suzuki generates ground state wave functions by setting up a grid and running time evolution given a set of initial conditions (configuration). This scales linearly with respect to the number of simulations. The advantage in using a trained GP as a predictive model is that multiple simulation runs can be batched together by converting different configurations $\{(x, g_1), (x, g_2), ... (x, g_M)\}$ into an input tensor of form $[[x, x, ... x], [g_1, g_2, ... g_M]]$. This trick makes it possible to pose multiple simulations as one prediction step. This is the reason for the significant speed up in generation of wave function using GPs compared to Trotter-suzuki. The empirical results discussed thus far indicate that Gaussian Processes are not just viable but desirable surrogate models for data-driven emulation of Bose-Einstein Condensates.

There are a number of avenues to explore which are beyond the scope of this work. We briefly mention these in this section. Vanilla Gaussian Processes do not scale well to high-dimensional data. modelling higher-dimensional BECs requires a scalable variant of GP. GP inherently models uncertainty in data. It is possible to exploit this in order to train GPs even more efficiently using Active Learning. We used different instances of GP to model wave functions conditioned on different trapping potentials. Using a common Multi-task GP (MTGP) \cite{mtgp} would increase the data-efficiency further as MTGP learns the common traits between wave functions generated in different settings.

\section{Conclusion}
\label{Conclusion}
Data-driven methods are replacing conventional numerical techniques in modelling physical phenomena in several fields of science. We explore data-driven emulation of BEC ground-state wave function using Gaussian Process as a surrogate model. Our empirical results show that Gaussian Processes in addition to modelling uncertainty from data, are able to surpass Artificial Neural Networks in terms of efficiency and complexity. Gaussian Processes when used as simulators post-training offer significant gain in speed when compared to Trotter-suzuki, a numerical approximation technique. Our experimental results indicate that GP is a desirable candidate for data-driven modelling of Bose-Einstein Condensates.

\section*{Acknowledgements}

R.Radha wishes to acknowledge financial assistance received from Council of Scientific and Industrial Research (No. 03(1456)/19/EMR-II Dated: 05/08/2019), Government of India.

\section*{Author contributions statement}

S.R. and T.B. conceived the experiments,  S.R. and T.B. conducted the experiments, M.S., R.R. and V.S. analysed the results.  All authors reviewed the manuscript. 

\section*{Additional information}

 \textbf{Competing interests:} 
 
The authors declare no competing interests.

\end{document}